\documentstyle[prb,epsfig,aps,amsmath,amssymb,multicol]{revtex}

\begin{document}
\draft

\title{Electrostatic Engineering of Nanotube Transistors
for Improved Performance}

\author{S.~Heinze$^*$, J.~Tersoff$^\dagger$, and Ph.~Avouris}

\address{IBM Research Division, T.~J.~Watson Research Center,
Yorktown Heights, New York 10598}

\date{\today}

\maketitle

\begin{abstract}
With decreasing device dimensions, the performance of carbon
nanotube field-effect transistors (CNFETs) is limited by high Off
currents except at low drain voltages. We show that an asymmetric
design improves the performance, reducing Off-currents and
extending the usable range of drain voltage. The improvement is
most dramatic for ambipolar Schottky-barrier CNFETs.  Moreover,
this approach allows a single device to exhibit equally good
performance as an n- or p-type transistor, by changing only the
sign of the drain voltage. Even for CNFETs having ohmic contacts,
an asymmetric design can greatly improve the performance for
small-bandgap nanotubes.
\end{abstract}

\pacs{}

\vspace*{-1cm}

\begin{multicols}{2}

\narrowtext

The performance of carbon nanotube field-effect transistors
(CNFETs) can be greatly improved in many respects by scaling the
devices to smaller
size~\cite{Dekker01,Heinze02,Appenzeller02,stefan-submitted}.
However, the improvements with smaller size are accompanied by the
undesired effect of a lowered On-Off ratio at typical drain
voltages. This restricts the range of usable drain voltage, which
in turn limits the achievable On-currents. Most CNFETs to date
operate as Schottky barrier (SB)
CNFETs~\cite{Dekker01,Heinze02,Appenzeller02,stefan-submitted,Martel01,Freitag01,MarkoAPL},
and the issue has been studied in detail for such
devices~\cite{MarkoAPL,Guo2003}. But the problem occurs even for
ohmic contacts, especially for small-bandgap tubes, as shown
below.

Here, we propose a novel device design for CNFETs that can be
scaled down in size for good turn-on performance without severe
restrictions on the usable drain voltage. The key idea of the new
design is to have large electric fields at the source contact but
small fields at the drain, to suppress unwanted tunneling. A
related approach has recently been demonstrated for Si-based
SB-FETs~\cite{Lin2001}.

For an ambipolar SB-CNFET, the asymmetric design can suppress
either the p- or the n-type branch of the ambipolar transport
characteristic, depending only on the {\it sign} of the drain
voltage. Thus the {\it same device} can act as an excellent p- or
n-type transistor, and the turn-on performance for both types is
controlled by the electrostatics (the geometry of the contact and
gate) at the source electrode. By choosing metals of different
work function for the gates, it should be possible to fabricate
complementary SB-CNFETs that have the desired alignment of
transfer characteristics ($I$ vs.~$V_{\rm g}$) for standard logic
applications.

For a SB-CNFET, the asymmetric design solves the problem of
Off-current that increases exponentially with drain voltage, so it
allows larger On-currents. This is a critical issue as the device
dimensions decrease~\cite{MarkoAPL,Guo2003}. Moreover, even for a
CNFET having ohmic contacts~\cite{DaiNature}, e.g.\ to the valence
band, suppressing the current into the conduction band can be
essential when the nanotube (NT) bandgap is small.

\begin{figure}
\begin{center}
\epsfig{file=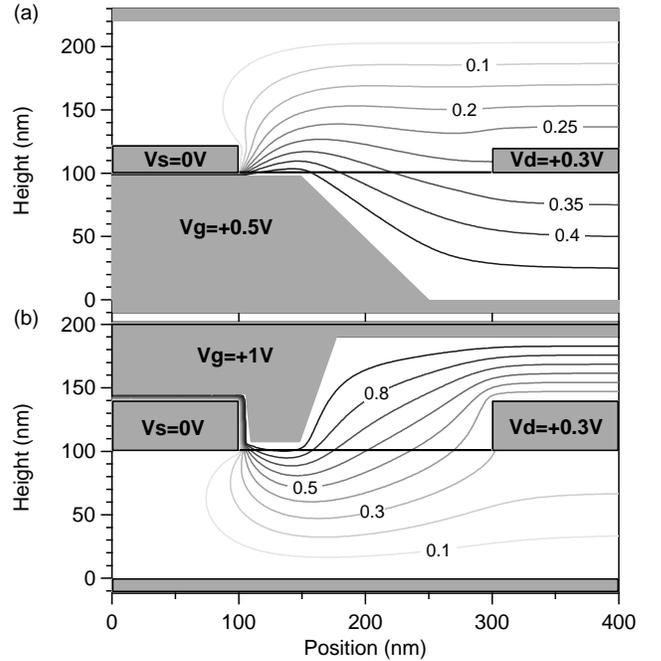,width=8.8cm}
\caption{\label{Fig:AsymmetricGeometries} Two possible geometries
for an asymmetric device. Contour lines show the electrostatic
potential. The applied voltages are given in the plot.  (a) A
bottom gate device with different oxide thicknesses at source
(2~nm) and drain (100~nm) contact. The top plane is grounded.
Adjacent contour lines differ by $0.05$~V.  (b) A local top gate
close to the source contact. The bottom plane is grounded.
Adjacent contour lines differ by $0.1$~V.}
\end{center}
\end{figure}

Two examples of possible asymmetric device geometries are shown in
Fig.~\ref{Fig:AsymmetricGeometries}. The first design,
Fig.~\ref{Fig:AsymmetricGeometries}(a), relies on having a
different thickness for the bottom oxide at the source and drain
contacts. The second design,
Fig.~\ref{Fig:AsymmetricGeometries}(b), employs a top gate close
to the source contact but not the drain. Other ways of engineering
the electric fields are possible, such as a sharp source electrode
to focus the field at the contact, and a blunt drain electrode to
reduce the field there.

To study the transport characteristics of these devices, we use a
semiclassical model which has been described
elsewhere~\cite{Heinze02,stefan-submitted,MarkoAPL}. We assume
ballistic transport within the NT (over the short distances of
interest here)~\cite{DaiNature,McEuen02}. For SB-CNFETs, the
current is controlled by the rate of thermally assisted tunneling
(at room temperature) through the SBs at the source and drain
contact.  We neglect charge on the NT, which is a good
approximation for the Off-state and in the turn-on
regime~\cite{Heinze02,Guo2003}.

For concreteness, we focus on the asymmetric bottom gate device,
Fig.~\ref{Fig:AsymmetricGeometries}(a). At the source contact the
oxide thickness, $t_{\rm ox}$, is only 2~nm. CNFETs with symmetric
source and drain contacts of such low oxide thickness have been
fabricated and show very good turn-on
performance~\cite{stefan-submitted,MarkoAPL}.
We first treat ideally ambipolar devices, where the metal Fermi
level falls in the middle of the NT bandgap at each contact. The
NT band gap is taken to be $0.6$~eV, corresponding to a diameter
of $1.4$~nm. Other cases are discussed below.

\begin{figure}
\begin{center}
\epsfig{file=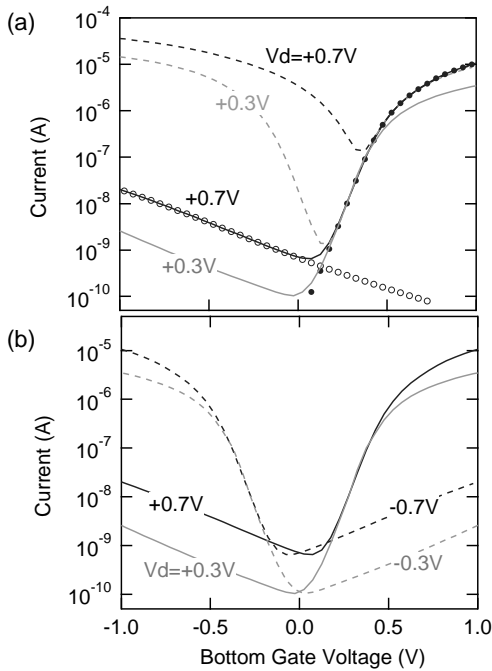,width=7.0cm}
\caption{\label{Fig:Current_vs_Vgate} (a) Calculated transfer
characteristics for a symmetric ($t_{\rm ox}=2$~nm, dashed lines)
and an asymmetric CNFET (geometry of
Fig.~\ref{Fig:AsymmetricGeometries}(a), solid lines). For the
asymmetric CNFET, the electron and hole contributions to the
current are given separately (filled and open circles,
respectively) for $V_{\rm d}=+0.7$~V. (b) Transfer characteristics
of the asymmetric CNFET at positive $V_{\rm d}$ (solid lines) and
negative $V_{\rm d}$ (dashed lines). }
\end{center}
\end{figure}

The calculated transfer characteristics for a {\it symmetric}
CNFET with $t_{\rm ox}=2$~nm are shown as dashed lines in
Fig.~\ref{Fig:Current_vs_Vgate}(a). The quality of the turn-on
performance can be quantified by the subthreshold slope, defined
by $S=(d\log_{10}{I}/dV_{\rm g})^{-1}$, where $I$ is the device
current and $V_{\rm g}$ is the gate voltage. The value of $S$ for
the symmetric CNFET of Fig.~\ref{Fig:Current_vs_Vgate}(a) is
$110$~mV/decade. (This is already within a factor of 2 of the
thermal limit of a conventional transistor, $\sim$60~mV/decade.)
However, the ratio between On- and Off-current decreases with
increased drain voltage $V_{\rm d}$. At $V_{\rm d}=+0.7$~V, the
ratio is only two orders of magnitude, well below the range
desired for transistors. This drain voltage effect has been
investigated in detail for such symmetric
CNFETs~\cite{MarkoAPL,Guo2003}, where it represents an important
limitation.

In contrast, the asymmetric design of
Fig.~\ref{Fig:AsymmetricGeometries}(a) allows a high On/Off ratio
even at higher $V_{\rm d}$, while preserving other good features
of the behavior. This is shown by the solid lines in
Fig.~\ref{Fig:Current_vs_Vgate}(a). At positive $V_{\rm d}$, the
slope in the turn-on regime (e.g.\  $V_{\rm g} \sim 0.3$) is the
same as for the symmetric device. However, the Off-current is much
lower than for the symmetric device,
and the On-Off ratio is greatly improved, to over 4 orders of
magnitude at $V_{\rm d}=+0.7$~V.

\begin{figure}
\begin{center}
\epsfig{file=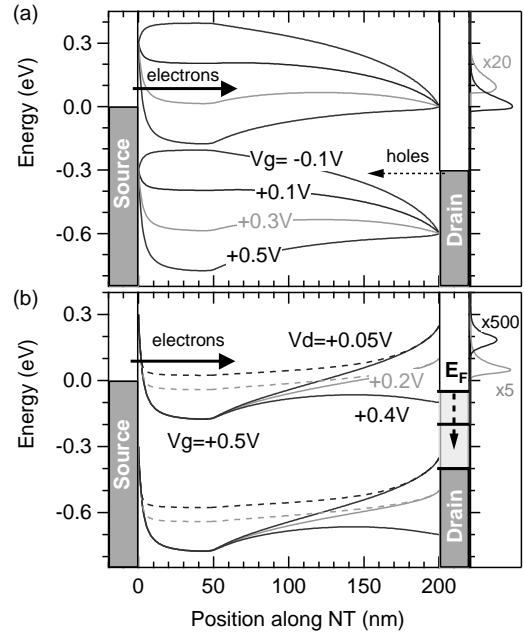,width=7.5cm}
\caption{\label{Fig:BandDiagrams} The calculated band bending
within the nanotube channel for the asymmetric device of
Fig.~\ref{Fig:AsymmetricGeometries}(a). (a)  Fixed $V_{\rm
d}=+0.3$~V with varying $V_{\rm g}$. The right side shows the
contribution to the current vs. energy in arbitrary units for
$V_{\rm g}=+0.3$~V (gray line) and $+0.5$~V (black line). (b)
Fixed $V_{\rm g}=+0.5$~V with varying $V_{\rm d}$. The right side
shows the contribution to the current vs. energy in arbitrary
units for $V_{\rm d}=+0.05$~V (black line) and $+0.2$~V (gray
line). Dashed lines illustrate (for $V_{\rm d}=+0.05$~V and
$0.2$~V) the effect of including charge on the NT, within a local
approximation~\cite{Heinze02}. While the band bending is
substantially different in the channel, the current in the
subthreshold regime is virtually unchanged~\cite{Footnote_Charge}.
}
\end{center}
\end{figure}

The contributions of electrons and holes to the current are shown
separately for one case in Fig.~\ref{Fig:Current_vs_Vgate}(a). For
the asymmetric CNFET at positive $V_{\rm d}$, the hole current is
suppressed, changing the behavior from ambipolar for a symmetric
device to n-type for the asymmetric device. By merely reversing
the {\it sign} of $V_{\rm d}$, one can instead suppress the
electron current, making the same asymmetric CNFET behave as a
p-type transistor with identical performance,
as shown in Fig.~\ref{Fig:Current_vs_Vgate}(b).

The transfer characteristics ($I$ vs.~$V_{\rm g}$)
of the asymmetric device can be understood based on the calculated
band bending shown in Fig.~\ref{Fig:BandDiagrams}(a) for $V_{\rm
d}=+0.3$~V. At large positive $V_{\rm g}$ ($+0.5$~V) the behavior
is similar to a symmetric device --- the SB at the source is
sufficiently thin for electrons to be injected, and there is
little or no barrier for them to be collected at the drain. Thus
the current is relatively large [see
Fig.~\ref{Fig:Current_vs_Vgate}(a)]. As $V_{\rm g}$ is reduced,
the source SB becomes wider, until electron injection becomes
negligible for $V_{\rm g} \lesssim +0.1$~V.

For negative $V_{\rm g}$, while the electron current remains
negligible, holes can be injected at the drain and are then easily
collected at the source. This is the origin of the ambipolar
characteristic (and of the undesirable Off-current) in symmetric
devices~\cite{MarkoAPL}. In our asymmetric device, however, the
geometry is chosen so that the SB at the drain remains rather
wide, suppressing hole tunneling. Thus the device shows purely
n-type character.

At negative $V_{\rm d}$, the argument is reversed --- electron
transport is suppressed due to the barrier at the drain contact,
while the hole current is as for a symmetric device. Thus by
simply changing the {\it sign} of $V_{\rm d}$, we can operate the
{\it same device} as either an excellent p- or n-type transistor,
with a turn-on performance controlled by the source contact
geometry. This is a key advantage of the new device design. In
principle, complementary n- and p-type transistors based on ohmic
contacts (to the conduction and valence band respectively) would
give even better performance; but the need for both kinds of
contacts presents a formidable obstacle to such a design at
present.

\begin{figure}
\begin{center}
\epsfig{file=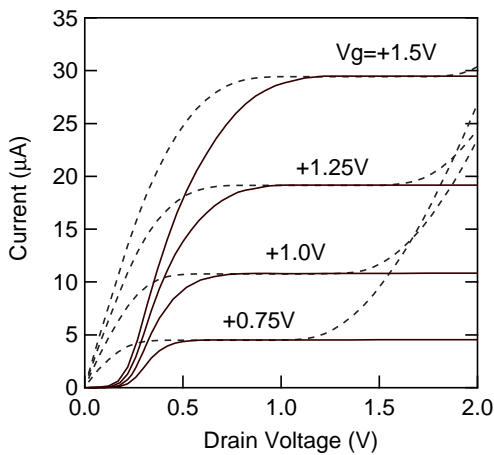,width=7.0cm}
\caption{\label{Fig:Current_vs_Vdrain} Calculated output
characteristics of the symmetric (dashed lines) and the asymmetric
(solid lines) CNFET. Device geometries are the same as for
Fig.~\ref{Fig:Current_vs_Vgate}.}
\end{center}
\end{figure}

The output characteristics ($I$ vs.~$V_{\rm d}$) of the asymmetric
and the symmetric CNFET are compared in
Fig.~\ref{Fig:Current_vs_Vdrain}. For the asymmetric device, the
current starts to rise appreciable only for $V_{\rm d} \gtrsim$
half the NT bandgap. The lack of a linear regime of $I$
vs.~$V_{\rm d}$ for the asymmetric device does not hinder
digital/logic applications which rely on the plateau of saturating
current. The plateaus are actually greatly improved for asymmetric
CNFETs. For symmetric devices, these plateaus are limited by the
onset of the drain leakage current~\cite{MarkoAPL}.

The output characteristics can be understood by examining the band
bending at fixed $V_{\rm g}$, Fig.~\ref{Fig:BandDiagrams}(b). At
$V_{\rm g}=0.5$~V, the SB at the source is sufficiently thin for
high transmission. However, at low $V_{\rm d}$ ($=+0.05$~V) the
electrons injected at the source contact cannot enter the drain
contact due to the thick potential barrier~\cite{Footnote_Charge}.
By increasing $V_{\rm d}$ the barrier at the drain contact is
lowered and the injected electrons can be collected at the drain.
Thus a $V_{\rm d}$ of about half the NT band gap is required for a
significant flow of current
(c.f.~Fig.~\ref{Fig:Current_vs_Vdrain}). With further increase of
$V_{\rm d}$ the current saturates, being limited by the source or
channel (both of which are controlled by $V_{\rm g}$ in the
absence of short-channel effects) rather than by the drain.

\begin{figure}
\begin{center}
\epsfig{file=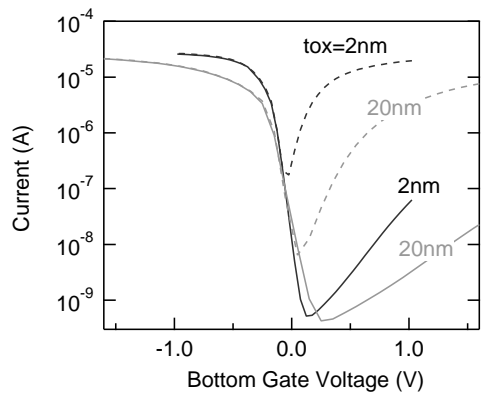,width=7.0cm}
\caption{\label{Fig:LowBandgap} Calculated transfer
characteristics at $V_{\rm d}=-0.2$~V for a NT with a bandgap of
0.3~eV and an ohmic contact to the valence band. Solid and dashed
lines are for asymmetric and symmetric CNFET, respectively.}
\end{center}
\end{figure}

So far we have focused on ambipolar SB-CNFETs; but ohmic contacts
could provide further performance improvements~\cite{DaiNature}.
Ohmic contacts are most readily achieved for NTs having small
bandgap~\cite{DaiNature}, in which case we find that drain leakage
current is still an issue, except at very small $V_{\rm d}$.
Figure~\ref{Fig:LowBandgap} shows calculations for a device having
ohmic contacts to the valence band, and where the NT bandgap is
0.3~eV, as in Ref.~\onlinecite{DaiNature}. As expected, the
subthreshold slope is excellent (65~mV/dec for $t_{\rm ox}=2$~nm).
However, for a symmetric CNFET, the SB to the {\it conduction
band} (i.e.\ the bandgap) is only $0.3$~eV.  Thus positive $V_{\rm
g}$ can induce substantial electron tunneling, leading to the
behavior shown by dashed lines in Fig.~\ref{Fig:LowBandgap}. This
electron tunneling is significant even for a 20~nm oxide; and for
very thin oxides the effect is dramatic, as shown for $t_{\rm
ox}=2$~nm.

For a CNFET having ohmic contacts, the leakage current can be
suppressed by an asymmetric geometry, just as for the ambipolar
devices described above. Thus, without excessively restricting
$V_{\rm d}$, one can achieve a dramatic improvement of the On-Off
ratio, suppressing the drain leakage current over a large range of
$V_{\rm g}$. As thin oxides are necessary for high performance
CNFETs, the asymmetric device design may play a key role even for
devices based on ohmic contacts, and even more so for Schottky
barrier transistors.

We thank Marko Radosavljevi\'{c} and Richard Martel for valuable 
discussions. S.\ H.\ thanks the Deutsche Forschungsgemeinschaft 
for financial support under the Grant number HE3292/2-1.

\end{multicols}

\end{document}